\documentclass[aps,prd,preprint,groupedaddress,tightenlines,floatfix,showkeys,11pt]{revtex4-2}

\usepackage{graphics}
\usepackage{graphicx}
\usepackage{amsmath,amssymb,empheq}
\usepackage{tabularx}
\usepackage{array}
\usepackage{parskip}

\begin{document}

\title{Photon escape from the ISCO of a rotating black hole in Asymptotic Safety}

\author{Miguel A. Enr\'\i quez}
\email{menriquezr@unal.edu.co}

\author{Luis A. S\'anchez}
\email{lasanche@unal.edu.co}

\affiliation{Departamento de F\'\i sica, Universidad Nacional de Colombia,
A.A. 3840, Medell\'\i n, Colombia}

%
\begin{abstract}
We study isotropic emission of photons from the innermost stable circular orbit (ISCO) of a subextremal 
rotating black hole (BH) in asymptotic safety (AS). We calculate both the photon escape probability (PEP) 
and the maximum observable blueshift (MOB) of photons to reach infinity, and compare with the corresponding 
results for photon emission from the ISCO of a classical Kerr BH. In AS, quantum gravity effects reduce the
ISCO radius, therefore quantum gravity effects should reduce the PEP and MOB of photons from emitters 
moving on the ISCO. We show that this is not the case and that, when rotating BHs with high spin are 
considered and the quantum parameter (which encodes the quantum gravity effects) increases towards its 
critical value, which is different for different spin values, the PEP and MOB also increase despite the 
reduction of the ISCO radius. Our results on the PEP show explicitly how quantum gravity effects start to dominate over the classical background at the ISCO scale. We also briefly discuss the relation between these quantum gravity modifications and particular features of the shadow of a rotating BH in AS.\\
\end{abstract}



\maketitle

\newpage
\section{\label{sec:intr}Introduction}
To date, general relativity (GR) has successfully passed all the experimental tests in the weak field 
regime. Recently, however, the detection of gravitational waves by the LIGO Collaboration \cite{ligo1,ligo2} 
and the observation of the shadow images of the supermassive BHs M87$^*$ and Sgr A$^*$ by the Event Horizon 
Telescope Collaboration \cite{EHT1,EHT2}, have opened the possibility to confront GR in the strong field 
regime of black hole (BH) horizon. In this concern, the important fact about the EHT images is that they are determined only 
by the space-time background metric in regions where high curvature effects become significant. 
This, in turn, provides the exciting opportunity to test possible deviations from GR arising from quantum 
theories of gravity, which aim to reconcile GR with quantum mechanics. In fact, the quantum gravity program 
is based on the belief that classical unphysical singularities associated to, e.g., the origin of the universe 
and those present inside the event horizon of black holes (BH), can be solved within the framework of a 
quantum theory of gravity.\\

\noindent
Asymptotic Safety (AS) is an approach to a quantum theory of gravity which is distinguished for being based entirely
on the standard quantum field theory as it makes use of the techniques of the functional renormalization
group (FRG) \cite{Reuter}. The FRG flow of the scale dependent quantum effective action for gravity is 
dictated by the Wetterich equation \cite{Wett}, which is a nonperturbative evolution equation. A theory 
is asymptotically safe if this flow approaches a non-Gaussian fixed point (NGFP) in the ultraviolet (UV) 
where quantum scale invariance is realized in the sense that a finite number of dimensionless coupling 
strengths become scale-invariant and safe from unphysical divergences. In AS the transition scale in the UV is assumed to be the Planck scale. Scale invariance in the UV ensures that the theory is 
UV-complete and predictive up to the highest energies \cite{Reut-Saue,Litim,Falls}. The onset of quantum 
scale invariance at the Planck scale leads to the antiscreening character 
of gravity in the UV, that is, to a weakening of gravity at high energies. The possibility of solving the problem of the singularity inside the horizon of black holes in the AS framework lies in this prediction\\
 
\noindent
To explore the consequences on the BH physics within the AS proposal for quantum gravity, the so-called renormalization group improved (RGI) metrics are constructed \cite{Bon-Reu}. These are obtained by replacing the classical metric with its scale-dependent counterpart that results from the RG flow. In the infrared (IR) limit, the improved metrics still depend on a free quantum parameter $\xi$, with the classical metrics reproduced in the 
limit $\xi \rightarrow 0$. So, the spacetime geometry we will consider is described by the IR or low energy limit of the RGI metric constructed in the context of an asymptotically safe theory of gravity containing high-derivative terms in the effective action \cite{Cai,Haroon}, which we will outline ahead. In this limit, the running Newton coupling acquires the form
\begin{equation} \label{GrCoup}
G(r) = G_0 \left(1-\frac{\xi G_0}{r^2}\right),
\end{equation}
where $G_{0}$ is the classical Newton constant and $\xi$ is the quantum parameter that determines the quantum
effects on the spacetime geometry. The impact of $\xi$ on the horizon scale physics of RGI BH 
spacetimes has been discussed in a plethora of works (for recent reviews see \cite{Eich,Plat} and 
references therein). In particular, quantum gravity modifications to the shadow cast by non-rotating 
and rotating BHs in AS have been studied in \cite{Eich2,LAS1,Shi}.\\

\noindent
It is well known that the morphology of the shadow of a spinning BH, that is, the size and asymmetry of 
the shadow and the brightness of the rings that bound it, is directly related both to the amount of 
photons that can escape to infinity and to the shift of their frequencies at infinity. In this direction, 
the problem of calculating the escape cone for monochromatic and isotropic photon emission from sources 
in the vicinity of the event horizon of a spherically symmetric BH has been studied in \cite{BaVi_1}, and for 
the Kerr BH was studied in \cite{BaVi_2,Semer}. The same study for the  
Kerr-de Stter BH and the connection with its shadow, was carried out in \cite{Stuch}. The photon 
escape probability (PEP) from a source in a local non-rotating frame in the vicinity of the horizon of 
a Kerr-Newman BH was analyzed in \cite{Ogas1}, and the same problem in the spacetime of a Kerr-Sen BH 
was studied in \cite{Zhang1,Zhang2}. The PEP and the maximum observable blueshift (MOB) of photons from emitters moving freely along geodesics on the equatorial plane near a Kerr BH, was considered in \cite{Gates,Su}. Then, in \cite{Igata1},
the authors investigate both the PEP and MOB of photons emitted from a light source falling from the 
innermost stable circular orbit due to an infinitesimal perturbation, and the effect of the 
proper motion of the source on the PEP and MOB. The PEP, when the emission point is not on the equatorial 
plane of a Kerr BH, was calculated in \cite{Ogas2,Ogas3}. The photon emission from equatorial stable 
and unstable circular orbits in the near-horizon extremal Kerr (NHEK) and near-horizon near-extremal 
Kerr (near-NHEK) regions has been investigated in \cite{Yan1,Yan2}. Also, the PEP and MOB of radiation 
emitted from the ISCO of a rapid rotating BH was analyzed in \cite{Igata2}.\\

\noindent
Our goal in this work is to study quantum gravity corrections to the PEP and MOB of photons emitted isotropically by an orbiter in geodesic motion on the ISCO of a rotating BH in AS, and compare our results with those already obtained for a Kerr BH in classical GR in \cite{Igata2}. We also look at the relation of these modifications to particular features of the shadow of the rotating RGI BH discussed in \cite{Eich2,LAS1}.\\

\noindent
This paper is organized as follows. The construction of the rotating RGI metric in the framework of the asymptotically safe gravity with high derivative terms, is presented in Sec.~\ref{sec:sec2}. In Sec.~\ref{sec:sec3} we study null geodesics in the RGI spacetime and the conditions for photon escape from the ISCO. Sec.~\ref{sec:sec4} is devoted to the dynamics of an orbiter traveling in a local reference frame moving along an equatorial geodesic orbit. In Sec.~\ref{sec:sec5} we construct the escape cone for a photon emitted from the source and calculate both the photon escape probability and the maximum observable blueshift. The connection between these results and the shadow cast by the RGI BH is discussed in Sec.~\ref{sec:sec6}. In the last section we summarize our results.
\section{\label{sec:sec2}Rotating black hole in asymptotic safety with higher derivatives}
In this section we outline the implementation of the IR limit of the running Newton coupling in Eq.~(\ref{GrCoup}) following Refs.~\cite{Cai,Haroon}. In these works the authors broadened the study of black hole solution in asymptotically safe gravity by including higher derivative terms in the effective action. In fact, they start with a generally covariant effective gravitational action given by
\begin{eqnarray}\nonumber
\Gamma_p[g_{\mu\nu}]&=&\int dx^4\sqrt{-g}~[p^4g_0(p)+p^2g_1(p)R +g_{2a}(p)R^2+g_{2b}(p)R_{\mu\nu}R^{\mu\nu} \\ \label{Action}
&+& g_{2c}(p)R_{\mu\nu\sigma\rho}R^{\mu\nu\sigma\rho}+ O(p^{-2}R^3)+ ...],
\end{eqnarray}
where $p$ is a momentum cutoff, $g$ is the determinant of the metric tensor $g_{\mu\nu}$, $R$ is the Ricci scalar, $R_{\mu\nu}$ is the Ricci tensor and $R_{\mu\nu\sigma\rho}$ is the Riemann tensor. The coefficients $g_{i} (i=0,1,2a,...)$ are dimensionless couplings satisfying the renormalization group equations
\begin{equation}\label{RG}
\frac{d}{d \ln p}g_{i}(p)=\beta_{i}[g(p)],
\end{equation}
where
\begin{equation}\label{ge01}
g_0(p) = -\frac{\Lambda(p)}{8 \pi G(p)} \, p^{-4}, \quad \quad
g_1(p) = \frac{1}{16\pi G(p)}\, p^{-2}\,.
\end{equation}
Here, $G(p)$ and $\Lambda(p)$ are the running Newton coupling and the running cosmological constant, respectively.\\
\noindent
With the aim of finding a vacuum static, spherically symmetric solutions to the theory, the following ansatz is proposed
\begin{equation}\label{ansatz}
ds^{2}=-f(r)dt^{2}+f(r)^{-1}dr^{2}+r^{2}d\Omega_2 ^{2}.
\end{equation}
After substituting this metric into the action (\ref{Action}) and solving the generalized Einstein field equations for vacuum
\begin{equation}\label{Einstein}
\tilde{G}^{\mu\nu}\equiv \frac{\delta\Gamma_p[g_{\mu\nu}]}{\delta g_{\mu\nu}}=0,
\end{equation}
a Schwarzschild-(anti)-de Sitter-like solution is obtained in the form
\begin{equation}\label{SAdS}
f(r)=1-\frac{2G(p) M}{r} - \frac{r^2}{l(p)^2},
\end{equation}
where $M$ is an integration constant that can be identified with the physical mass of the BH, and $l(p)$ is the radius of the asymptotic (anti)-de Sitter space, respectively.\\
\noindent
It is convenient to rewrite the coefficients of the high-derivative terms in the action in terms of a new set of coefficients $(\tilde{\theta}, \tilde{\omega}, \tilde{\lambda}$), in the form
\begin{equation}\label{ges}
g_{2a} = -\frac{1}{6\tilde{\lambda}}+\frac{\tilde{\theta}}{\tilde{\lambda}}+\frac{\tilde{\omega}}{3\tilde{\lambda}}~,\quad \quad
g_{2b} = \frac{1}{\tilde{\lambda}}-\frac{4\tilde{\theta}}{\tilde{\lambda}}~,\quad \quad
g_{2c} = -\frac{1}{2\tilde{\lambda}}+\frac{\tilde{\theta}}{\tilde{\lambda}}~.
\end{equation}
In this way, the coefficient $\tilde{\lambda}$ can be written in the familiar logarithmic form as in asymptotic freedom
\begin{equation}\label{lambda}
\tilde{\lambda}(p) =\frac{\tilde{\lambda}_0}{1+\frac{133}{160\pi^2}\tilde{\lambda}_0\ln (p/M_p)},
\end{equation}
with $\tilde{\lambda}_0$ being a fixed value of $\tilde{\lambda}$ at the Planck scale $M_p$.\\

\noindent
Asymptotic safety requires that all the beta functions vanish, $\beta_i[g(p)]=0$, when the dimensionless couplings $g_i$ approach a fixed point $g_{i}^*$. Then, solving the beta functions for the gravitational coupling and the cosmological constant, an IR Gaussian fixed point and an UV non-Gaussian fixed point are obtained. They arise in the form
\begin{eqnarray}\label{IRg0}
g_0&\simeq&-\frac{({\Lambda}_0+\gamma{p^2}G_0)(1+\zeta{p^2}G_0)}{8\pi{p}^4G_0}~,\\ \label{IRg1}
g_1&\simeq&\frac{1+\zeta{p^2}G_0}{16\pi{p}^2G_0}~,
\end{eqnarray}
where $G_0$ and ${\Lambda}_0$ are the IR values of the gravitational coupling (the Newton constant) and the cosmological constant. The coefficients $\gamma$ and $\zeta$ are small free parameters which capture the quantum effects on the spacetime
geometry. From Eqs.~(\ref{ge01}) and (\ref{IRg1}) we get for the running Newton coupling
\begin{eqnarray}\label{Gp}
G(p)=\frac{1}{16\pi{p}^2g_1}=\frac{G_0}{1+\zeta{p^2}G_0}~.
\end{eqnarray}
Interestingly, $G(p)$ decreases for increasing values of the cutoff $p$ which implies a weakening of gravity at high energy scales, while at low energy scales ($p \rightarrow 0$) coincides with the Newton constant $G_0$.\\
\noindent
The values of $g_0$ and $g_1$ at the UV non-Gaussian fixed point have been obtained by numerically solving the flow equation. The result is \cite{Cai}
\begin{equation}\label{gesUV}
g_0^* \simeq -\frac{\gamma\zeta G_0}{8\pi} \simeq-6.331\times10^{-3}~,\quad \quad
g_1^* \simeq \frac{\zeta}{16\pi} \simeq1.432\times10^{-2}~.
\end{equation}
Since our aim in the present study is to examine quantum gravity effects on the PEP and MOB of photons emitted by an orbiter moving on the ISCO radius, we must determine the scale identification between the momentum cutoff $p$ and the radial coordinate $r$ in the low energy limit, that is, for $r >> l_{Planck}$. In this limit, the high-derivative terms are suppressed and we can establish the known identification $p \sim 1/r$. Then, Eq.~(\ref{Gp}) becomes
\begin{eqnarray}\label{Gr}
G(r)\simeq G_0 \left(1-\frac{\xi G_0}{r^2} \right)~,
\end{eqnarray}
where $\xi$ is a constant which deviates from the coefficient $\zeta$ by a factor of $\mathcal{O}(1)$ and where an expansion at leading order in $\xi$ has been done.\\

\noindent
Cosmological implications associated to the running of $\Lambda(p)$ will not be considered in this work. In the remainder of this paper we adopt geometrized units: $G_{0} = c = 1$.\\

\noindent
Replacing Eq.~(\ref{Gr}) into (\ref{SAdS}) and ignoring the cosmological constant term, the static spherically symmetric line element acquires the Schwarzschild-like form
\begin{equation}\label{Schw}
ds^2=-\left(1-\frac{2{\cal M}(r,\xi)}{r}\right)dt^2+\left(1-\frac{2{\cal M}(r,\xi)}{r}\right)^{-1}dr^2+r^2 
d\Omega^2,
\end{equation}
where the coupling in Eq.~(\ref{Gr}) has been translated into a Misner-Sharp mass function given by
\begin{equation}
{\cal M}(r,\xi) = M\Big(1-\frac{\xi}{r^2}\Big),
\label{Meff}
\end{equation}
with $M$ being the mass of the BH. Note that the classical Schwarzschild spacetime is retrieved in the 
limit $\xi\rightarrow{0}$.\\

\noindent
The associated rotating BH metric is obtained by applying the generalized Newman-Janis algorithm \cite{NJ} (see also \cite{LAS1}). Thus, the Kerr-like metric involving the running Newton coupling in Eq.~(\ref{Gr}) can be written, in 
Boyer-Lindquist coordinates, as
\begin{eqnarray}\nonumber
ds^2&=&-\Big(1-\frac{2{\cal M}(r,\xi) r}{\Sigma}\Big)dt^2-\frac{4a{\cal M}(r,\xi) r\sin^2\theta}{\Sigma}dtd\phi+\frac{\Sigma}{\Delta}dr^2+\Sigma{d\theta^2}\\\label{ASmetric}
&+&\sin^2\theta\bigg(r^2+a^2+\frac{2a^2 {\cal M}(r,\xi) r}{\Sigma}\sin^2\theta\bigg)d\phi^2,
\end{eqnarray}
where $a$ is interpreted as the BH spin and
\begin{align}
\Delta=r^2-2{\cal M}(r,\xi)r+a^2,
\quad \quad
\Sigma=r^2+a^2 \cos^2{\theta}.
\end{align}
Also in this case must be noticed that the classical Kerr BH is recovered in the 
limit $\xi\rightarrow{0}$.

\begin{figure}
 \includegraphics[width=.50\linewidth]{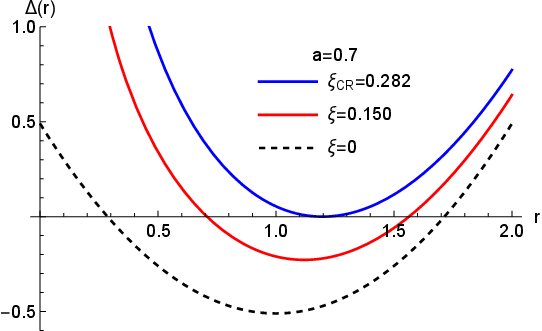} \quad \quad
  \includegraphics[width=.40\linewidth]{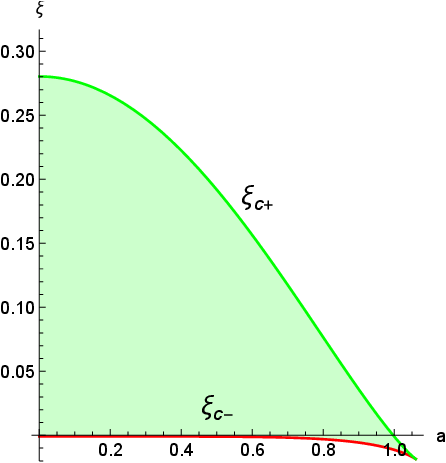}
\caption{Left panel: the function $\Delta(r)$ as a function of the radial coordinate for $a^{\star}=0.7$ and $\xi=\xi_{\rm CR}\simeq 0.282$ (blue), $\xi=0.150$ (red), and for $\xi=0$ which corresponds to the classical BH (black dashed). 
Right panel: plots of the parameters $\tilde{\xi}_{c+}$ (green) and $\tilde{\xi}_{c-}$ (red) as functions of the spin parameter $a$. The green region is the parameter space for allowed BH solutions.}
\label{horizons}
\end{figure}

\noindent
From now on we fix $M=1$, so that all quantities are expressed in units of the BH mass: $r/M \rightarrow r$, $\xi/M^{2} \rightarrow \xi$, $a/M \rightarrow a$, etc. Then, the condition $\Delta=0$ that determines the new horizons, reads
\begin{equation}
r^3 - 2r^{2}+{a}^{2} r+2 \xi=0.
\label{Delta}
\end{equation}
The discriminant of this cubic equation is
\begin{equation}
D_3=-{a}^6+{a}^4+(16-18 {a}^2)\xi-27 \xi^2,
\label{discrim}
\end{equation}
and the condition $D_3=0$ is a quadratic equation in $\xi$ solved by
\begin{equation}
\xi_{c\pm} = \frac{\pm\sqrt{\left(4 -3 {a}^{2}\right)^3}-9 {a}^{2} + 8}{27}.
\label{xipm}
\end{equation}
It follows that real values of $\xi$ in the range $\xi_{c-}\leq \xi < \xi_{c+}$ 
require $4\geq 3 {a}^{2}$ and that, as shown in Ref.~\cite{LAS2} and as in the Kerr solution, there are
two nonzero horizons only for $0\leq \xi <\xi_{c+}$, where $\xi_{c+}\equiv \xi_{\rm CR}$ is the critical 
value for which the two horizons merge as shown in Fig.~\ref{horizons} (left panel). From the right 
panel of Fig.~\ref{horizons} we see that the extreme Kerr BH with $a=1$ coincides with the quantum corrected 
BH with $a=1$ and $\xi=0$, and that, at difference with the classical BH, for any value of the spin
($0\leq a \leq 1$), and for $\xi=\xi_{c+}$ the quantum BH is extremal. Moreover, there exist a real 
solution for $a=2/\sqrt{3}>1$ and $\xi=-4/27$ with no classical analogue. It is important to note that the outer horizon of the RGI BH shrinks as $\xi$ increases, that is, as the quantum gravity effects become significant.\\ 

\noindent
In what follows, we will consider values of $\xi$ in the interval $0\leq \xi <\xi_{\rm CR}$, which means that we will study photon escape from the ISCO of a non-extremal RGI BH. 
%
\section{\label{sec:sec3}Photon emission}
The metric in Eq.~(\ref{ASmetric}), being stationary and axisymmetric, has two Killing vectors and, 
consequently, two associated conserved quantities: the energy $E$ and the axial component of the 
angular momentum $L_z$. There is one additional conserved quantity associated to the Killing-Yano tensor
field: the Carter constant ${\cal Q}$. These three quantities are related to each other through
the quadratic integral of motion ${\cal K}$ as follows \cite{CHANDRA}
\begin{equation}
{\cal Q} = {\cal K} - (L_z-aE)^2.
\label{Carter}
\end{equation}
Rescaling the photon four-moment as $k^{\mu}/E\rightarrow k^{\mu}$, the null-geodesic equations are
\begin{align}\label{null1}
k^t& = \dot{t} =\frac{1}{\Sigma}\Big( -a^2 \sin^2\theta+ \frac{1}{\Delta}[\left(r^2+a^2)^2-2a\eta r{\cal M}(r,\xi)\right]\Big),\\ \label{null2}
k^r& =  \dot{r} = \sigma_r \frac{\sqrt{{\cal R}(r)}}{\Sigma},\\ \label{null3}
k^{\theta}& = \dot{\theta} = \sigma_{\theta} \frac{\sqrt{\Theta(\theta)}}{\Sigma},\\ \label{null4}
k^{\phi}& = \dot{\phi} = \frac{1}{\Sigma}\Big(\frac{\eta^2}{\sin^2\theta}+\frac{a}{\Delta}\left[2r{\cal M}(r,\xi)-a\eta\right]\Big), 
\end{align}
where the dots denote derivative respect to an affine parameter along the geodesic, $\sigma_r=\pm$, 
$\sigma_{\theta}=\pm$, and
\begin{eqnarray}\label{Erre}
{\cal R}(r)&=& r^4+\left(a^2-\eta^2-\chi\right)r^2+2r\left[\chi+(\eta-a)^2\right]{\cal M}(r,\xi)-a^2\chi,\\ \label{Teta}
\Theta(\theta)&=& \chi +a^2\cos^2\theta-\eta^2\cot^2\theta. 
\end{eqnarray}
Here, the two energy-rescaled conserved parameters $\chi$ and $\eta$ are defined as
\begin{equation}\label{Impact}
\chi = \frac{{\cal Q}}{E^2}, \quad \quad \eta = \frac{L_z}{E}.
\end{equation}
To study the photon escape from the ISCO to infinity, we restrict ourselves to the equatorial plane ($\theta=\pi/2$) and introduce the effective potentials $\eta_1$ and $\eta_2$ by writing the radial equation (\ref{null2}) in the form
\begin{equation}\label{EcDif}
\dot{r}^2 + \frac{r(r-2 {\cal M})}{\Sigma ^2}(\eta-\eta_1)(\eta-\eta_2)=0,
\end{equation}
where
\begin{eqnarray}\label{eta1}
\eta_1(r,\xi) & = & \frac{-2 a {\cal M} r + \sqrt{r \Delta \left[r^3- (r-2 {\cal M})\chi \right]}}{r (r-2 {\cal M})},\\ \label{eta2}
\eta_2(r,\xi) & = & \frac{-2 a {\cal M} r-\sqrt{r \Delta \left[r^3- (r-2 {\cal M})\chi \right]}}{r (r-2 {\cal M})}.
\end{eqnarray}
\noindent
The extrema of the effective potentials correspond to the radii $r^{\rm cl,q}_1$ and $r^{\rm cl,q}_2$ ($r^{\rm cl,q}_1 < r^{\rm cl,q}_2$) of the unstable spherical photon orbits (SPO), where ``$\textrm{cl}$'' and ``$\textrm{q}$'' stand for ``classical'' and ``quantum'', respectively. The expressions for the conserved impact parameters at these radii are given by the conditions ${\cal R}(r)=0$ and $\partial{\cal R}(r)/\partial r=0$, which provide \cite{DIMNI}
\begin{eqnarray}\nonumber
\chi_{\rm SPO}(r,\xi) &=& \frac{r^3}{a^2(r-{\cal M}-r{\cal M}^{\prime})^2}\biggl[4a^2{\cal M}-r(r-3{\cal M})^2\\ \label{chi}
&-& 2r{\cal M}^{\prime}
(r^2-3r{\cal M}+2a^2)-r^3({\cal M}^{\prime})^2 \biggr], \\ \label{eta}
\eta_{\rm SPO}(r,\xi) &=& \frac{(r^2-a^2){\cal M}-r\Delta -r(r^2+a^2){\cal M}^{\prime}}{a(r-{\cal M}-r{\cal M}^{\prime})},
\end{eqnarray}
where the prime denotes derivative with respect to $r$.\\
\begin{figure}[t!]
 \includegraphics[width=.55\linewidth]{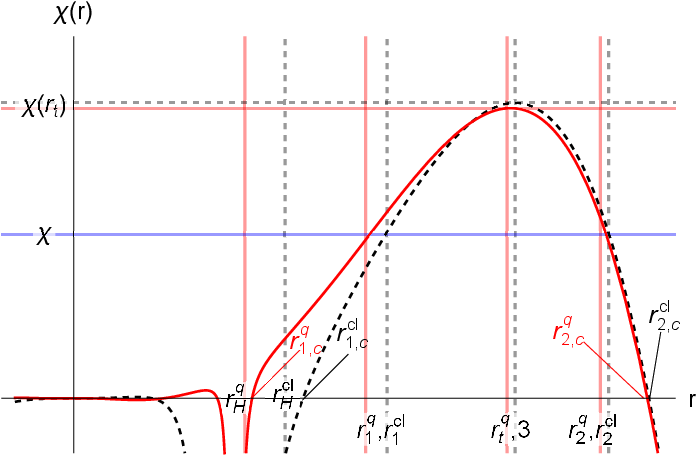} \quad \quad
\caption{The impact parameter $\chi^{\rm cl}_{\rm SPO}(r,0)$ for a Kerr BH (black dashed) and $\chi^{\rm q}_{\rm SPO}(r,\xi=0.07)$ for a RGI BH (red) for $a=0.9$ The vertical red and black dashed lines locate the positions of the couples of radial coordinates that are relevant to describe the photon escape. The horizontal blue line shows the relation between a fixed value of $\chi^{\rm cl,q}_{\rm SPO}$ and the radii $r^{\rm cl,q}_i$ ($i=1,2$) of the SPO}
\label{CHI}
\end{figure}

\noindent
Let's call $\chi(r,0)\equiv \chi^{\rm cl}$ and $\chi(r,\xi \neq 0)\equiv \chi^{\rm q}$. Fig.~\ref{CHI} shows plots of $\chi^{\rm cl,q}_{\rm SPO}(r,\xi)$ for a subextremal classical (black dashed curve) and a subextremal RGI (red continuous curve) black hole, both with $a=0.9$. For the AS BH we take 
$\xi=0.07<\xi_{\rm CR}\simeq 0.099$. 
The plots show only the region $\chi^{\rm cl,q}_{\rm SPO}(r,\xi) \geq 0$, a restriction that comes from Eqs.~(\ref{null3}) and (\ref{Teta}) because a particle moving in the equatorial plane must have $\chi\geq 0$. We see that the curve for the RGI BH is shifted to the left with respect to the classical case and, consequently, all the relevant radial coordinates are shifted to smaller values. We also see that the AS curve expands in width but contracts in height such that the classical maximum at $r=3$ ($\chi^{\rm cl}_{\rm SPO}=27$) changes to $r_{\rm t}<3$ ($\chi^{\rm q}_{\rm SPO}(r_{\rm t})<27$). The horizontal blue line shows the relation between a fixed value of $\chi^{\rm cl,q}_{\rm SPO}$ and the radii $r^{\rm cl,q}_i$ ($i=1,2$) of the SPO. They are in the range determined by the two roots $r^{\rm cl,q}_{\rm 1,c }$ and $r^{\rm cl,q}_{\rm 2,c}$ of the equations 
$\chi^{\rm cl,q}_{\rm SPO}(r,\xi)=0$ which define the radii of circular photon orbits in the equatorial plane, i.e., $r^{\rm cl,q}_{\rm 1,c} < r^{\rm cl,q}_1 < r^{\rm cl,q}_2 < r^{\rm cl,q}_{\rm 2,c}$. In the classical theory, solving the equation $\chi^{\rm cl}_{\rm SPO}(r,0)=0$, the radii of circular photon orbits in the equatorial plane are obtained as
\begin{align}
&r^{\rm cl}_{1,\mathrm{c}}=2+2\cos \left[\frac{2}{3} \arccos(a)-\frac{2\pi}{3}\right],
\\
&r^{\rm cl}_{2,\mathrm{c}}=2+2\cos \left[\frac{2}{3} \arccos(a)\right].
\end{align}
In the AS theory, these radii must be calculated numerically.\\

\begin{figure}[t!]
 \includegraphics[width=.48\linewidth]{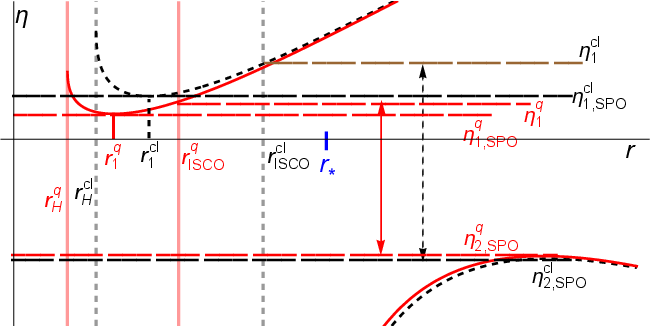}
  \includegraphics[width=.48\linewidth]{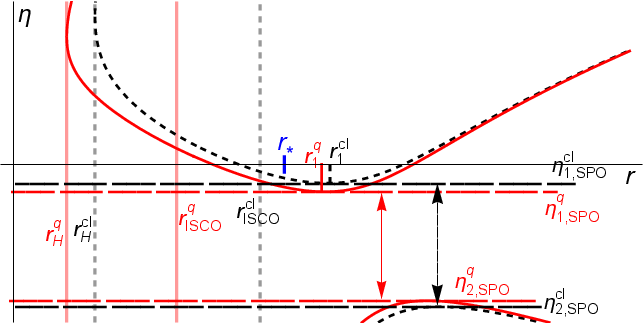}
\caption{Plots of the effective potentials $\eta^{\rm cl,q}_1$ and $\eta^{\rm cl,q}_2$ for $a=0.9$. The black dashed curves are for the classical Kerr BH, while the red curves are for the RGI BH with $\xi=0.07<\xi_{\rm CR}\simeq 0.099$. The impact parameter $\chi$ has been set to $\chi=10$ (left panel) and $\chi=25$ (right panel). Left panel defines case (a): $r^{\rm cl,q}_H < r^{\rm cl,q}_1 < r_{\star}$. Right panel defines case (b) $r^{\rm cl,q}_H < r_{\star} < r^{\rm cl,q}_1$.}
\label{ETAS}
\end{figure}

\noindent
In Fig.~\ref{ETAS}, we show typical plots of $\eta^{\rm cl,q}_1$ and $\eta^{\rm cl,q}_2$ for the same values of $a$ and $\xi$ as in Fig.~\ref{CHI} and two fixed values of $\chi$: $\chi=10$ (left panel) and $\chi=25$ (right panel). Again, and in accordance with what Fig.~\ref{CHI} shows, the quantum plots are shifted to the left with respect to the classical plots and thus all the relevant radial coordinates satisfy: $r^{\rm q}_{\rm H} < r^{\rm cl}_{\rm H}$ (the horizon radius), $r^{\rm q}_{\rm ISCO} < r^{\rm cl}_{\rm ISCO}$ (the ISCO radius), and $r^{\rm q}_i < r^{\rm cl}_i$ ($i=1,2$). It is important to highlight that the horizon radius and the ISCO radius are closer to each other for the RGI BH than they are for the Kerr BH.\\ 

\noindent
Since the escape of photons from a BH is dictated by the effective potentials $\eta^{\rm cl,q}_1$ and 
$\eta^{\rm cl,q}_2$, the information provided by Figs.~\ref{CHI} and \ref{ETAS} allow us to establish the conditions for photon escape in terms of the set of parameters $(\sigma_r,\chi^{\rm cl,q},\eta_i^{\rm cl,q})$. From Fig.~\ref{ETAS} we can identify two cases depending of the location of the source respect the horizon radius an the extreme $r^{\rm cl,q}_1$ of $\eta^{\rm cl,q}_1$: 
\begin{itemize}
\item (a) Right panel: $r^{\rm cl,q}_H < r^{\rm cl,q}_1 < r_{\star}$. In this case, for values of 
$\chi^{\rm cl,q}_{\rm SPO}(r,\xi)$ in the range $0\leq \chi^{\rm cl,q}_{\rm SPO}(r,\xi)<\chi^{\rm cl,q}_{\rm SPO}(r_{\star},\xi)$, we see that: (i) photons emitted from the ISCO with outward radial momentum 
($\sigma_r=+$) can escape from the BH for values of $\eta^{\rm cl,q}$ satisfying: 
$\eta^{\rm cl,q}_{2, \rm SPO}< \eta^{\rm cl,q}\leq \eta^{\rm cl,q}_1$, i.e., in the range between the horizontal black dashed lines marked by the vertical black dashed double arrow (classical Kerr BH), and in the range between the horizontal red dashed lines marked by the vertical red double arrow (RGI BH); (ii) photons emitted from $r_{\star} \geq r^{\rm cl,q}_{\rm ISCO}$ with inward radial momentum ($\sigma_r=-$) can escape from the BH when $\eta^{\rm cl,q}_{1, \rm SPO}<\eta^{\rm cl,q}<\eta^{\rm cl,q}_1$, i.e., for $\eta^{\rm cl,q}$ in the range between the two upper horizontal black dashed lines (classical case) and in the range bounded by the two upper horizontal red dashed lines (quantum gravity case). Thus, the so called marginal values of 
$\eta^{\rm cl,q}$ are identified as $(\sigma_r,\eta^{\rm cl,q})=(+,\eta^{\rm cl,q}_{2, \rm SPO})$ and  
$(\sigma_r,\eta^{\rm cl,q})=(-,\eta^{\rm cl,q}_{1, \rm SPO})$. We note that, interestingly, photons with inward momentum which are captured by the classical BH, can escape from the AS BH, and that the region for escaping photons with $\sigma_r=+$, narrows.
\item (b) Left panel: $r^{\rm cl,q}_H < r_{\star} < r^{\rm cl,q}_1$. Only photons with outward radial momentum 
($\sigma_r=+$) can escape from the BH and so, for $\eta^{\rm cl,q}$ in the range $\eta^{\rm cl,q}_{2, \rm SPO}< \eta^{\rm cl,q}\leq \eta^{\rm cl,q}_{1, \rm SPO}$. The classical range is identified by the vertical black double arrow, and by the vertical red double arrow for the RGI BH. The marginal values of 
$\eta^{\rm cl,q}$ are then identified as $(\sigma_r,\eta^{\rm cl,q})=(+,\eta^{\rm cl,q}_{2, \rm SPO})$ and  
$(\sigma_r,\eta^{\rm cl,q})=(+,\eta^{\rm cl,q}_{1, \rm SPO})$. We see that in this case more photons can reach infinity from the classical Kerr BH that from the BH in AS.
\end{itemize}
\section{\label{sec:sec4}Circular equatorial orbiter and stationary local frame}
For a particle following a stable geodesic orbit in the equatorial plane, the specific 
energy $\kappa$ and specific angular momentum $\lambda$, that is, the energy and angular momentum per
unit rest mass of the particle in prograde motion, are given by \cite{LAS2}
\begin{eqnarray}\label{Espe}
\kappa=\frac{r^3-2(r^2-\xi)+ a \sqrt{r\left(r^2-3 \xi \right)}}{r^{3/2}\sqrt{r^3-3r^2+5\xi + 2 a \sqrt{r\left(r^2-3 \xi \right)}}},
\end{eqnarray}
\begin{eqnarray}\label{Lspe}
\lambda=\frac{(r^2+a^2)\sqrt{r\left(r^2-3 \xi \right)} - 2a(r^2-\xi)}{r^{3/2}\sqrt{r^3-3r^2+5\xi + 2a \sqrt{r\left(r^2-3 \xi \right)}}}.
\end{eqnarray}
For these orbits the angular velocity acquires the form
\begin{equation}\label{omega}
\Omega=\frac{\left(r^2-3 \xi \right)^{1/2}}{{r^{5/2} + a}\left(r^2-3 \xi \right)^{1/2}},
\end{equation}
and the effective potential per unit mass $V_\text{eff}(r)$, which can be read from the radial equation of motion:
\begin{eqnarray}
\frac{1}{2}\dot{r}^2+V_\text{eff}(r)=\frac{1}{2}\left(\kappa^2-1\right),\label{rad}
\end{eqnarray}
is given by
\begin{equation}\label{Veff}
V_{\rm eff}(r)=-\frac{{\cal M}(r,\xi)}{r}+\frac{\lambda^2-a^2(\kappa^2-1)}{2r^2}-\frac{{\cal M}(r,\xi)(\lambda-a\kappa)^2}{r^3}
\end{equation}
The radius $r_{\rm ISCO}$ of the innermost stable circular orbit occurs at the local minimum of the effective potential, so it is calculated from
\begin{eqnarray}\label{isco}
\frac{d^2V_\text{eff}}{dr^2}|_{r=r_{\rm ISCO}}=0.
\end{eqnarray}
For the classical Kerr BH ($\xi=0$), the solution for co-rotating orbits is
\begin{equation}\label{iscoCL}
r^{\rm cl}_{\rm ISCO}=3 + Z_2 - \sqrt{(3-Z_1)(3+Z_1+2Z_2)},  
\end{equation}
where
\begin{eqnarray}\nonumber
Z_1 &=&1 + \sqrt[3]{1-a^2}(\sqrt[3]{1-a}+\sqrt[3]{1+a}), \\ \nonumber
Z_2 &=& \sqrt{3 a^2 + Z_1^2} \nonumber
\end{eqnarray}
In the AS theory, as in the case for the radii of circular photon orbits, $r^{\rm q}_{\rm ISCO}$ must be obtained numerically.\\

\noindent
To describe physics in a local frame, we choose the stationary local tetrad as the one tied to the particle (source) in geodesic motion along constant $r$ and with $\theta=\pi/2$ (freely orbiting observers \cite{Semer2}), that is
\begin{align}\label{et}
e^{(t)}&=-\kappa \:\!dt+\lambda \:\!d\phi,
\\ \label{er}
e^{(r)}&=\frac{r}{\sqrt{\Delta}}\:\!dr,
\\ \label{etheta}
e^{(\theta)}&=r\:\!d\theta,
\\  \label{ephi}
e^{(\phi)}&=-\frac{\sqrt{\Delta } r^{3/2}}{\delta_1}\kappa \:\!dt+\frac{\sqrt{\Delta } \sqrt{r(r^2-3 \xi) }}{\delta_2\:\!\Omega }\lambda\:\!d\phi,
\end{align}
where
\begin{eqnarray}\label{del1}
\delta_1&=& a \sqrt{r} \sqrt{r^2-3 \xi }+r^3-2 \left(r^2-\xi \right),\\ \label{del2}
\delta_2&=& \sqrt{r} \left(a^2+r^2\right) \sqrt{r^2-3 \xi }-2 a \left(r^2-\xi \right)
\end{eqnarray}
For $\xi=0$, Eqs.~(\ref{et}-\ref{ephi}) coincide with the corresponding ones given in Refs.~\cite{Igata2} and \cite{Semer2}.\\
\noindent
The components of the photon four-momentum in the local frame are calculated as $k^{(a)}=k^{\mu}e^{(a)}_{\mu}$, where the $k^{\mu}$ are given by Eqs.~(\ref{null1}-\ref{null4}) and the $e^{(a)}_{\mu}$ come from Eqs.~(\ref{et}-\ref{ephi}). Then, we have
\begin{eqnarray}\nonumber
k^{(t)}&=& \frac{1}{r^2} \biggl[\left(\eta -a+\frac{a \left(a^2-a \eta +r^2\right)}{\Delta}\right)\lambda \\ \label{kt}
&-& \left(a (\eta -a)+\frac{\left(a^2+r^2\right) \left(a^2-a \eta +r^2\right)}{\Delta}\right)\kappa \biggr],\\ \label{kr}
k^{(r)} &=& \sigma_r\frac{1}{r} \sqrt{\frac{R}{\Delta}}, \\ \label{ktheta}
k^{(\theta)} &=& \sigma_{\theta}\frac{\sqrt{\chi }}{r}, \\ \nonumber
k^{(\phi)}&=& \frac{1}{\sqrt{\Delta} r^{3/2}}\biggl[-\frac{r \left(\left(a^2+r^2\right) \left(a^2-a \eta +r^2\right)+a \Delta (\eta -a)\right)}{\delta_1}\kappa \\ \label{kphi}
&+& \frac{\sqrt{r^2-3 \xi} \left(a \left(a^2-a \eta +r^2\right)+\Delta (\eta -a)\right)}{\delta_2 
\Omega}\lambda \biggr]
\end{eqnarray}
We will use these quantities to calculate the PEP an MOB of radiation emitted from the ISCO of the RGI rotating BH.
%
\section{\label{sec:sec5}Photon escape probability and maximum observable blueshift}
In this section, with the aim of confront our results with those of the classical theory, we closely follow Ref.~\cite{Igata2} where the authors study the PEP and MOB for photon emission from the ISCO of a Kerr BH.
\begin{figure}[b!]
 \includegraphics[width=.30\linewidth]{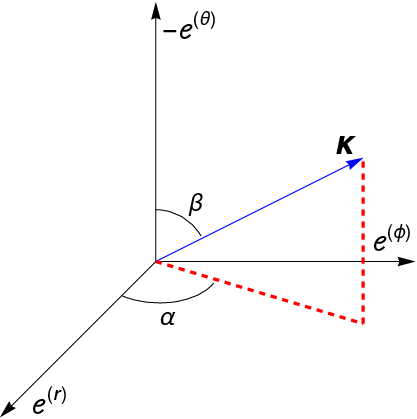} \quad \quad
\caption{Parameterization of the orbiter sky by the local angles ($\alpha,\beta)$.}
\label{AngDef}
\end{figure}
\subsection{\label{sec:sec5.1} Photon escape probability}
We start by parametrizing the sky of the emitter in geodesic equatorial motion using the local angles $(\alpha, \beta)$ as depicted in Fig.~\ref{AngDef}, where $\alpha\in (-\pi,\pi)$ is the azimuthal angle measured from the $e^{(r)}$ direction towards the $e^{(\phi)}$ direction in the plane expanded by $(e^{(r)},e^{(\phi)})$, and $\beta\in (0,\pi)$ is the polar angle from the local zenith that is, from the $-e^{(\theta)}$ direction. So we write
\begin{alignat}{3}\nonumber
\sin \alpha=\frac{k^{(\phi)}}{\sqrt{(k^{(r)})^2+(k^{(\phi)})^2}},
\quad&&
\cos \alpha&=\frac{k^{(r)}}{\sqrt{(k^{(r)})^2+(k^{(\phi)})^2}},\\ \nonumber
\\
\sin \beta=\frac{\sqrt{(k^{(r)})^2+(k^{(\phi)})^2}}{k^{(t)}}, \label{Angles}
\quad&&
\cos \beta&=-\frac{k^{(\theta)}}{k^{(t)}}.
\end{alignat}
From the discussion on the photon escape regions in Sec.~\ref{sec:sec3} (see Figs.~\ref{CHI} and \ref{ETAS}), we can realize that the angles $(\alpha, \beta)$ are functions of $\sigma_r$, $\eta^{\rm cl,q}$, $\chi^{\rm cl,q}_{\rm SPO}$ and $r_{\star}$, which we write as $\big(\alpha,\beta\big)(\sigma_r,\eta^{\rm cl,q},\chi^{\rm cl,q}_{\rm SPO},r_{\star})$, and that the the marginal values of $\eta^{\rm cl,q}$ imply that the critical emission angles $(\alpha, \beta)$ for photons escaping from the BH with $r_{\star} \geq r^{\rm cl,q}_{\rm ISCO}$, are given by

\begin{itemize}
\item $\big(\alpha_1, \beta_1\big)\equiv \big(\alpha,\beta\big)(-,\eta^{\textrm{cl,q}}_{1,\textrm{SPO}})$
\quad for \quad $0_{-}<\chi^{\textrm{cl,q}}_{\rm SPO}<\chi^{\textrm{cl,q}}_{\textrm{SPO}}(r_{\star})$,
\item $\big(\alpha_2, \beta_2\big)\equiv \big(\alpha,\beta\big)(+,\eta^{\textrm{cl,q}}_{1,\textrm{SPO}})$ 
\quad for \quad $\chi^{\textrm{cl,q}}_{\textrm{SPO}}(r_{\star})\leq \chi^{\textrm{cl,q}}_{\rm SPO}<\chi^{\textrm{cl,q}}_{\rm SPO}(r_t)$,
\item $\big(\alpha_3, \beta_3\big)\equiv \big(\alpha,\beta\big)(+,\eta^{\textrm{cl,q}}_{2,\textrm{SPO}})$ 
\quad for \quad $\chi^{\textrm{cl,q}}(r_{\textrm{t}})\geq \chi^{\textrm{cl,q}}_{\rm SPO} \geq 0_{+}$,
\end{itemize}
where $0_{-}$ is the zero of $\chi^{\rm cl,q}_{\rm SPO} $ at $r^{\rm cl,q}_{1,\rm c}$, and $0_{+}$ is the zero of 
$\chi^{\rm cl,q}_{\rm SPO}$ at $r^{\rm cl,q}_{2,\rm c}$, and, as mentioned above, $r_{\rm t}=3$ and 
$\chi^{\rm cl}_{\rm SPO}(3)=27$ for a Kerr BH.\\
\begin{figure}[t!]
    \includegraphics[width=0.45\linewidth]{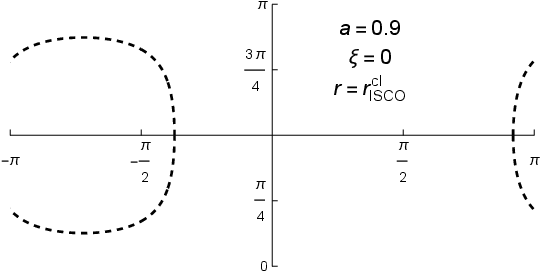}\\
    \includegraphics[width=0.45\linewidth]{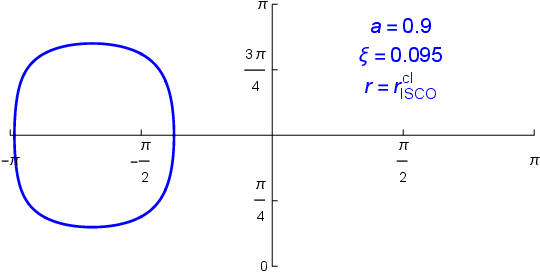}\\
    \includegraphics[width=0.45\linewidth]{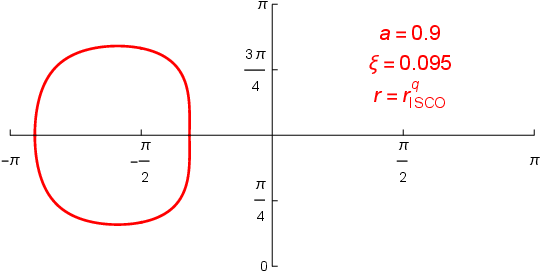}
\caption{Escape cone for a photon emitted from a source moving on the ISCO of a subextremal BH with spin parameter $a=0.9$. The upper panel is for a Kerr BH, the middle panel is for a RGI BH with the source moving on the classical ISCO, and the bottom panel is for a RGI BH with the orbiter moving along the ISCO calculated from the AS theory. The escape cone corresponds to the region bounded by the curves in the $(\alpha,\beta)$ plane that contains the coordinate origin.} \label{EscCone}
\end{figure}

\noindent
In Fig.~\ref{EscCone} we show plots of the critical angles for emission from a source moving on the ISCO of a subextremal BH with a high spin parameter $a=0.9$. The escape cone corresponds to the region bounded by the curves in the $(\alpha,\beta)$ plane that contains the coordinate origin \cite{Igata2}. The upper panel is for the classical Kerr BH ($\xi=0$) for which $r^{\rm cl}_{\rm ISCO}\simeq 2.32$. In this case our plot coincides with the one obtained in Ref.~\cite{Igata2} for the same value of $a$. The middle panel is for a RGI BH with $\xi=0.095 < \xi_{\rm CR}\simeq 0.099$ and, as in the upper panel, for emission from a source moving on the classical ISCO. The lower panel is for the same AS BH as in the middle panel but with the source moving on the ISCO of the RGI BH which takes the value $r^{\rm q}_{\rm ISCO}\simeq 1.488$ for $\xi=0.095$. The remarkable fact about these plots is that, contrary to what happens in the classical GR theory where the area of the escape cone reduces as the radius of the emission point becomes smaller, in the AS theory a smaller orbiter radius $r_{\star}$ leads to an increase of this area. Clearly, we have a genuine quantum effect associated to the weakening of gravity at high curvature scales predicted by the AS theory. This quantum effect should lead to an increase of the photon emission probability (see the next section) as $r_{\star}$ approaches the ISCO of the RGI BH, which is also the opposite of what happens in the classical theory.\\
\begin{figure}[t!]
    \includegraphics[width=1\linewidth]{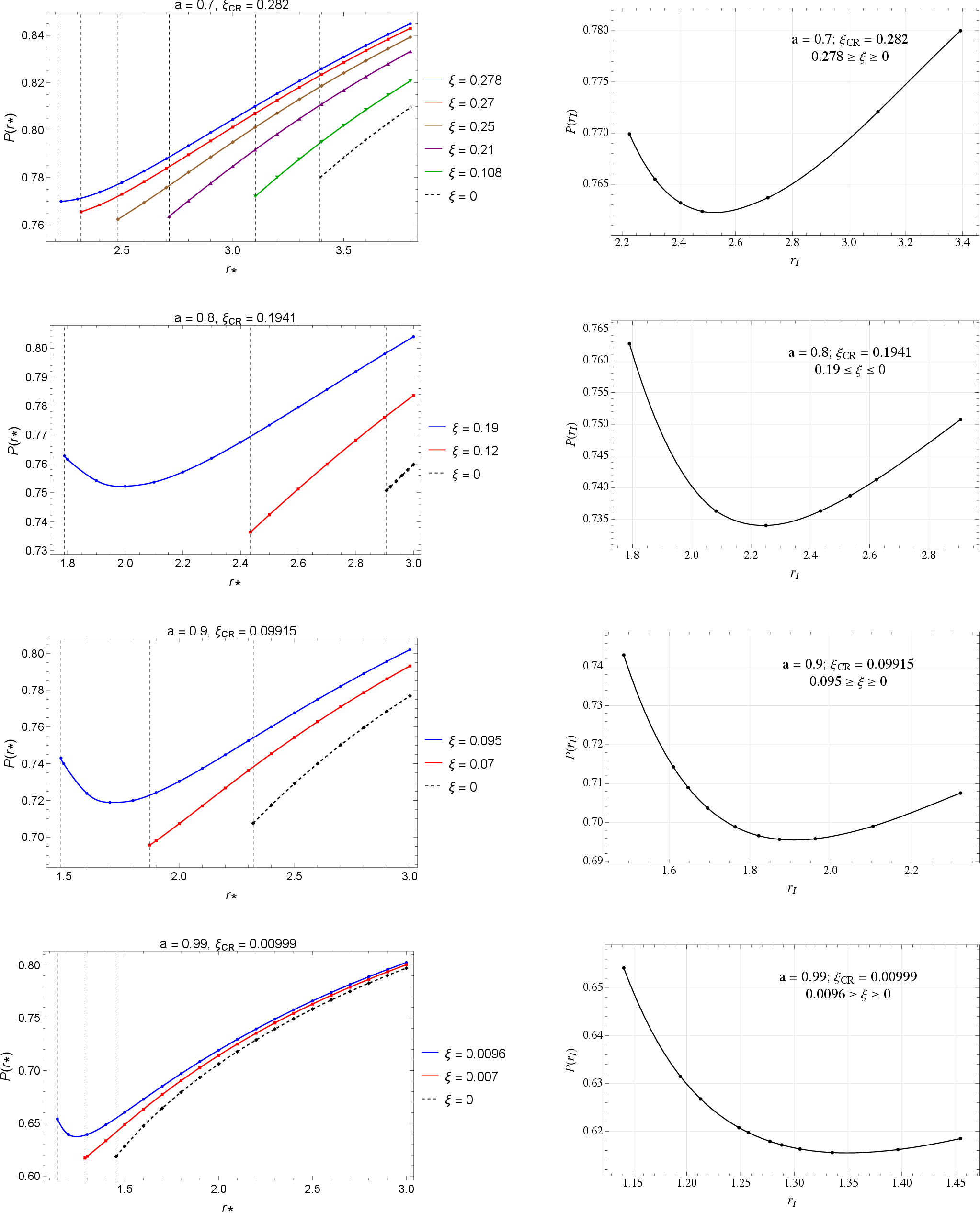}
\caption{Photon escape probability (P) versus position of the source ($r_\star$). Left column: $P$ vs.  $r_{\star}$ for increasing values of $a$ (from top to bottom). Within each plot $\xi$ grows from bottom to top with $\xi < \xi_{\rm CR}$. The black dashed curve is the PEP for a Kerr BH. The vertical dashed lines locate the ISCO for each value of 
$\xi$. Right column: $P$ vs. $r_{\rm ISCO}$. Each plot is the envelope of the corresponding curves on the left with more data added.} \label{PROB}
\end{figure}
\noindent
Now, assuming that photon emission is isotropic in the orbiter frame, the escape 
probability of a photon escaping to infinity $P$ is defined as the fraction of solid angle in the emitter sky corresponding to the escape directions, i.e.,
\begin{align}
P=\frac{{\cal{A}}_e}{4 \pi}=\frac{1}{4\pi}\int_S d\alpha\:\!d \beta \sin \beta =-\frac{1}{4\pi}\int_S d\alpha\:\! d\cos \beta,
\end{align}
where ${\cal{A}}_e$ is the area of the escape region in the emitter sky and $S$ is the escape cone.
In terms of critical angles and with ${\cal{A}}_c=4\pi - {\cal{A}}_e$ being the area of the capture region,
the escape probability can be written as 
\begin{eqnarray}\nonumber
P(r_{\star})&=& 1-\frac{{\cal{A}}_c}{4 \pi} \\ \nonumber
&=& 1- \frac{1}{2\pi} \int_{r_{1}^\mathrm{c}}^{r_{\star}} d r
\frac{d \alpha_{1}}{d r}\cos \beta_{1}
- \frac{1}{2\pi} \int_{r_{\star}}^{r_t} d r 
\frac{d \alpha_{2}}{d r}\cos \beta_{2}\\
& & -\frac{1}{2\pi} \int_{r_t}^{r_{2}^{\mathrm{c}}} d r 
\frac{d \alpha_3}{d r}\cos \beta_3,
\end{eqnarray}
where we have taking into account the prescription for the critical emission angles discussed above.\\

\noindent
Fig.~\ref{PROB} shows our results for the PEP as a function of $r_{\star}$ for different values of the spin parameter $a$ of a subextremal BH. The values of $a$ grow from the top panel ($a=0.70$) to the bottom one ($a=0.99$). Each panel in the left column shows the values of the PEP for emission from a source located at $r_{\star}\geq r_{\rm ISCO}$ with the ISCO radius in the horizontal axis identified by a vertical dashed line. The black dashed curve in each panel corresponds to the PEP for a Kerr BH for which our results are in accordance with the values reported in \cite{Igata2}. The colored curves are plots of the PEP for the BH in the AS theory, with the values of the quantum parameter $\xi$ growing from bottom to top within each panel, as shown. Clearly, the ISCO radius decreases as $\xi$ increases towards its critical value where the RGI BH becomes extremal (see Fig.~\ref{horizons}). The panels in the right column show the PEP from the ISCO as a function of $r_{\rm ISCO}$, so these curves are the envelope of the corresponding curves on the left with more data added. Interestingly, from the right side of the left panels it can be observed that, for $r_{\star}\geq r^{\rm cl}_{\rm ISCO}$, the PEP for emission from the RGI BH is always greater than the PEP for emission from the Kerr BH. Even more interestingly we observe that, unlike what happens with the classic curves, as $r_{\star}$ decreases towards $r^q_{\rm ISCO}$, the curves for the RGI BH develop an inflection point and they smoothly bend upward in such a way that with the increase of $a$ and $\xi$, the bending becomes progressively more pronounced with respect to the value of the PEP at $r^{\rm cl}_{\rm ISCO}$. Note also that for moderate values of the spin parameter ($a\lesssim 0.7$), the change in concavity becomes less noticeable, and that, for a fixed value of $a$ ($a\gtrsim 0.7$) and growing values of $\xi$, the value of the PEP at $r^{\rm q}_{\rm ISCO}$ for the RGI BH exceeds the value of the PEP at $r^{\rm cl}_{\rm ISCO}$. As referred to in the previous section, the behavior of the enveloping curves when $r_{\star}\rightarrow r^{\rm q}_{\rm ISCO}$, is a consequence of the antiscreening character of gravity at short distances predicted by the AS theory. In this sense, the inversion of concavity explicitly shows the tension between classical and quantum gravity effects in the vicinity of the horizon of a RGI BH.
%
\begin{figure}[b]
 \includegraphics[width=.40\linewidth]{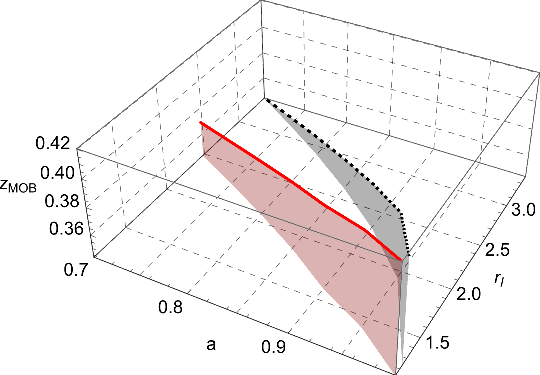} \quad \quad
\caption{Plots of $z_{\rm MOB}$ as a function of $a$ and $r_{\star}=r_{\rm ISCO}$. The black dashed curve is for a Kerr BH. The red curve is for the AS BH with fixed value of $\xi\lesssim \xi_{\rm CR}$ for each value of the spin parameter $a$.}
\label{MOBS}
\end{figure}
\subsection{\label{sec:sec5.2} Maximum observable blueshift}
In terms of the energy-rescaled photon four-momentum, the redshift factor $g$ and blueshift factor $z$ of a photon reaching the infinity with energy $E$ are defined, respectively, by
\begin{equation}\label{MOB}
g=\frac{1}{-k^{(t)}}, \quad \quad z=1-\frac{1}{g}=1+k^{(t)}.
\end{equation}
It can be shown that, in the classical case and for $r_{\star}\geq r^{\rm cl}_{1,\rm c}$, $z$ takes its maximum value $z_{\rm MOB}$ for $\chi^{\rm cl}_{\rm SPO}=0$, $\eta_1^{\rm cl}(r=r_{\star})$ and $\sigma_r=+$ \cite{Igata1}. Obviously, the same prescription is valid for the RGI BH. Fig.~\ref{MOBS} shows the behaviour of the values of $z_{\rm MOB}$ as a function of $a$ and $r_{\star}=r_{\rm ISCO}$. The black dashed curve is for a Kerr BH, while the red curve is for the AS BH with fixed value of $\xi\lesssim \xi_{\rm CR}$ for each value of $a$. We find that for all values of ($a,r^{\rm cl}_{\rm ISCO}$) and ($a,r^{\rm q}_{\rm ISCO}$), $z_{\rm MOB}$ is greater for the RGI BH than for the Kerr BH except for $a=1$ when both coincide. Our results for $z_{\rm MOB}$ as a function of $r_{\star}$ for the Kerr BH are the same than the obtained in \cite{Su}. It is important to point out that the increase in $z_{\rm MOB}$ is not related to an increase of the orbital velocity of the source due to the shrinking of the ISCO. In fact, as shown in \cite{LAS2}, the values of the angular momentum of a massive particle moving freely in the equatorial plane around a RGI BH, are displaced toward smaller values as $r_\star \rightarrow r^{\rm q}_{\rm ISCO}$. Thereby, precisely because 
$r^{\rm q}_{\rm ISCO}<r^{\rm cl}_{\rm ISCO}$ for all values of $\xi$ ($0\leq \xi \lesssim \xi_{\rm CR}$), a greater $z_{\rm MOB}$ is also attributable to the weakening of gravity in the vicinity of the horizon because a weaker gravitational pull allows Doppler blueshift due to relativistic beaming overcomes the gravitational redshift more efficiently.

\section{\label{sec:sec6}PEP, MOB and the Shadow of the RGI BH}
\begin{figure}[h!]
 \includegraphics[width=.45\linewidth]{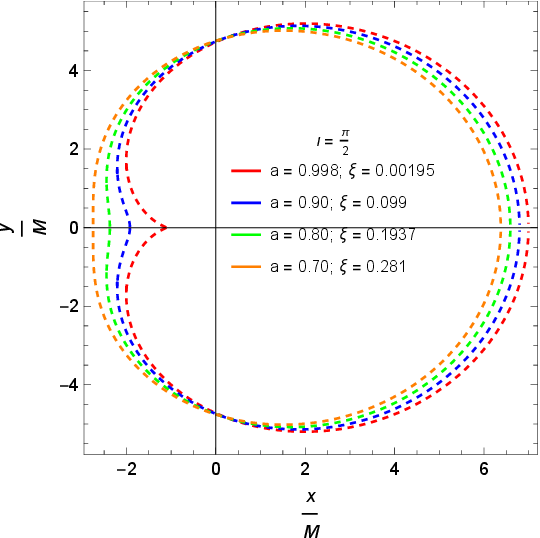} \quad \quad
\caption{Shadows cast by the spinning RGI BH for the observer inclination
angle $\iota=\pi/2$. Each shadow silhouette is constructed for the
values ($a,\xi\lesssim \xi_{\rm CR}$)) as shown.}
\label{SHAD}
\end{figure}

In Fig.~\ref{SHAD} we reproduce some of the plots we obtained in Ref.~\cite{LAS1} for the shadow contour of a RGI BH for selected values of $(a, \xi \lesssim \xi_{\rm CR})$ and for an observer inclination angle $\iota=\pi/2$. Let's remember that $\iota$ is defined as the angle between the
line of sight of the observer and the rotation axis of the BH. We see that for $a\gtrsim 0.8$, just at the left end of the contour which corresponds to photons in prograde orbits, appears
a kind of cusp-like effect. Therefore, since the ISCO of a RGI BH shrinks compared to that of a Kerr BH, the frame-dragging effect is stronger for prograde photons which leads to the distortion of the shadow of the rotating RGI BH compared to the typical D-shaped structure of the shadow of the classical Kerr BH. This means that for RGI BHs, prograde photons probe regions where the quantum gravity effects manifest themselves with greater intensity. Moreover, when $a\gtrsim 0.7$, $\xi \gtrsim \xi_{\rm CR}$, and 
$r_\star \rightarrow r^q_{\rm ISCO}$, we are just in the region where the PEP increases even though $r_\star$ decreases, which is also a quantum gravity effect. Since the brightness around the shadow depends on the PEP, that is, on how often photons can escape from the orbiter to infinity, and since its observability depends on the magnitude of the MOB, the photon rings that define the edge of the shadow of a spinning RGI BH should be brighter and more easily observable than expected, particularly at the end where photons are in prograde motion.\\
\section{\label{sec:sec7}Summary and final remarks}
This work has focused on the calculation of the photon escape probability (PEP) and the maximum observable blueshitf (MOB) of photons emitted from a source moving on the ISCO of a subextremal rotating BH in asymptotic safety, as well as on the comparison with the corresponding results obtained in the classical theory of GR. We have also briefly discussed the connection of our results to the deformation of the shadow of a RGI BH induced by quantum gravity effects. The spacetime geometry we consider is described by the infrared limit of a renormalization group improved rotating metric. In this case, in addition to the usual parameters $M$ and $a$, the black hole is also characterized by a free parameter $\xi$ that describes the quantum gravity effects on the geometry. Our analysis is rooted in the standard description of equatorial motion in Kerr spacetime, so that our calculation scheme of null and timelike geodesics, orbital motion, local frames, and photon propagation has been formulated in a geometrically transparent way.\\

\noindent
Upon introducing the RGI metric, the structure of the impact parameter $\chi$, the effective potentials
$\eta_i$ ($i=1,2$) for photons, and geodesic motion of massive particles, are modified. In fact we have found that all the relevant coordinates for studying the dynamics of the source and the emitted photons, namely: the horizon radius, the ISCO radius, the radii of the SPO, and the radii of circular equatorial orbits, are shifted to smaller values, thus giving rise to quantitative corrections in the conditions determining marginal photon trajectories for photon escape. The first indication we have found of significant modifications has come from the fact that, for relatively high values of the spin parameter ($a\gtrsim 0.7$) and a value of $\xi$ growing towards its critical value for each value of $a$, the area of the photon escape cone grows even though the radial location of the source $r_\star$ decreases towards $r_{\rm ISCO}$. The immediate consequence for the RGI BH is the increase of the PEP for decreasing locations of the source, a behavior opposite to that known in GR. The modification is even more dramatic since the plots of the PEP versus the location of the emitter, develop an inflection point such that they undergo a change in concavity that smoothly bends the curve upwards (see Fig.~\ref{PROB}). Surprisingly, for high but no critical values of $a$ and $\xi$, the bending allows for greater values of the PEP when evaluated at the ISCO of the RGI BH than the values of the PEP at the ISCO of the Kerr BH, despite $r^{\rm q}_{\rm ISCO}$ is always smaller than $r^{\rm cl}_{\rm ISCO}$. Although this behavior is seemingly counterintuitive, can be explained in the context of AS theory as a consequence of the prediction of weakening of the gravitational interaction at the horizon scale where the spacetime curvatue is considerable high. In this sense it is interesting to note how the inversion in concavity of curves in Fig.~\ref{PROB} explicitly shows the tension between classical and quantum gravity effects in the vicinity of the RGI BH horizon.\\

\noindent
Next, we found that the MOB for photons emitted from an orbiter at the ISCO of the RGI BH is higher than the MOB from an emitter at the ISCO of a Kerr BH. This is also a consequence of the antiscreening character of gravity in the strong field regime. In this case however, a greater MOB is not the result of an increase of the orbital velocity of the source due to the shrinking of the ISCO. Indeed, it has been shown in \cite{LAS2} that the values of the angular momentum of a massive particle in geodesic equatorial motion around a RGI BH are displaced toward smaller values as $r_\star \rightarrow r^{\rm q}_{\rm ISCO}$. So, higher values of the MOB occur because a weaker gravitational pull allows Doppler blueshift due to relativistic beaming overcomes the gravitational redshift more efficiently.\\

\noindent
Finally, we have drawn attention to the fact that the region of the parameter space ($a,\xi$) for which a cusp-like structure arises in the shadow silhouette of the RGI BH at the end corresponding
to photons in prograde orbits (see Refs.~\cite{Eich2} and \cite{LAS1}), just coincides with the region of the parameter space for which the increase of the PEP starts to become more conspicuous. This fact, along with the increase of the MOB for photons emitted from the ISCO of a RGI BH, leads us to argue that, since the brightness around the shadow depends on the magnitude of the PEP, and since the observability of horizon scale physics should be favored for greater MOB values, the photon rings that define the edge of the shadows of a spinning RGI BH should be brighter and more easily observable by a distant observer than expected, particularly at the end where photons are in prograde motion. We must emphasize that our findings should not be exclusive of the AS theory. Cusp-like structures are also found in the BH shadow of models of effective quantum gravity \cite{Ban}, in the Konoplya-Zhidenko rotating BH \cite{Wei}, and even in the rotating traversable wormhole \cite{Cheng}. In the same token, increase of the PEP at the horizon scale for extreme and non-extreme Kerr-Sen black hole has been reported in \cite{Zhang1, Zhang2}.\\

\noindent
Obviously, the present work can be extended to the study of different types of photon emitters moving in the neighborhood of a RGI BH. For example, the analysis of photon emission from plunging emitters will enable evaluation of quantum gravity modifications to high-energy emission from near horizon accretion flows. This would be of great relevance to future astrophysical observations in the near horizon regime that could allow detection of deviations from the predictions of GR. As shown in detail in \cite{Fosch}, it should be more likely to find these deviations in stellar-mass BHs (StMBH) which, at present, are the smallest known singularities. For near-extreme rotating StMBHs, the horizon radius and a prograde ISCO can be very close, therefore quantum gravity induced effects should have a greater chance to be observed in these astrophysical objects.\\

\noindent
Overall, our results indicate that AS inspired quantum gravity corrections preserve the
robustness of classical Kerr phenomenology for moderate values of the spin parameter $a$, but become
dynamically relevant in the strong field regime for high values of $a$ and near-critical values of $\xi$. In that limit, the increasing proximity of the ISCO to the horizon (see Fig.~\ref{ETAS}) acts as a geometric amplifier, converting local modifications arising from the RGI gravitational coupling into potentially observable optical effects. The PEP and the MOB therefore emerge as potential indirect probes of the effective quantum structure of spacetime in the strongest accessible gravitational regime. 

\section*{ACKNOWLEDGEMENTS}
We thank Takahisa Igata for useful and insightful comments. We acknowledge financial support from Universidad Nacional de Colombia.

\end{document}